\documentstyle[twoside,fleqn,espcrc2,epsf]{article}

\newcommand{\AmS}{{\protect\the\textfont2
  A\kern-.1667em\lower.5ex\hbox{M}\kern-.125emS}}

\title{\vspace{-1cm}\begin{flushright}
{\small UTCCP-P-76 \\
}
\end{flushright}
Heavy-light spectrum and decay constant from NRQCD with two
flavors of dynamical quarks\thanks{talk presented by A.~Ali~Khan}}

\author{
{CP-PACS Collaboration } :
A.~Ali~Khan\address{Center for Computational Physics,
University of Tsukuba, Tsukuba, Ibaraki 305-8577, Japan}, 
S.~Aoki\address{Institute of Physics, 
University of Tsukuba, Tsukuba, Ibaraki 305-8571, Japan},
R.~Burkhalter$^{\rm a,b}$,
S.~Ejiri$^{\rm b}$, 
M.~Fukugita\address{Institute for Cosmic Ray Research, 
University of Tokyo, Tanashi, Tokyo 188-8502, Japan}, 
S.~Hashimoto\address{High Energy Accelerator Research Organization (KEK), 
Tsukuba, Ibaraki 305-0801, Japan}, 
N.~Ishizuka$^{\rm b}$, 
Y.~Iwasaki$^{\rm a,b}$, 
K.~Kanaya$^{\rm a,b}$, 
T.~Kaneko$^{\rm a}$, 
Y.~Kuramashi$^{\rm d}$,
T.~Manke$^{\rm a}$,
K.~Nagai$^{\rm a}$,
M.~Okawa$^{\rm d}$, 
H.P.~Shanahan\address{DAMTP, 21 Silver St., University of Cambridge,
Cambridge, CB3 9EW, England, U.K.},
A.~Ukawa$^{\rm a,b}$ , and T.~Yoshi\'e$^{\rm a,b}$
}

\begin{document}
\begin{abstract}
We report on a study of $B$ mesons on $N_f = 2$ full QCD configurations 
using an RG-improved gauge action, 
NRQCD heavy quark action  and tadpole-improved clover light quark action. 
Results on the heavy-light spectrum and the decay constants
from $16^3\times 32$ lattices at $a^{-1}\approx 1.5$ GeV are presented, 
and compared with quenched results obtained with the same action combination
at matching lattice spacings.
\end{abstract}

\maketitle

\section{Introduction}
The decay constant $f_B$ is being studied extensively on the lattice
because of its importance for the determination of CKM matrix elements.
The spectrum of excited $B$ mesons and $b$ baryons is being measured in
present experiments, whereas there exist only few lattice results on this
subject.

In this article we report on our study of $B$ mesons in two-flavor full 
QCD employing the NRQCD action for heavy quark and 
a tadpole-improved clover action for light quark.  The dynamical 
configurations have been generated using the same light quark action and 
an RG-improved gauge action with a plaquette and a rectangular term. 
Details on our full QCD configurations can be found 
in Refs.~\cite{cp-pacs-full}.
A parallel study of $B$ mesons using the clover action for heavy quark 
is presented in Ref.~\cite{hugh}.

\section{Simulation Details}
We present results for two sets of dynamical lattices 
corresponding to the heaviest and the lightest sea quark in our 
configuration set at $\beta=1.95$.  
The results are compared to those from quenched lattices generated with 
the same RG-improved gauge action at $\beta=2.187$, the
lattice spacing from the string tension 
matched to the dynamical lattice with $\kappa_{sea} = 0.1375$. 
Some details on these runs are given in Table~\ref{tab:params}.

\begin{table}[t]
\caption{Parameters of lattices.  
The statistics for the dynamical lattices has been increased since 
Lattice'99. The scale is fixed by $\sqrt{\sigma}=427$ MeV 
(for each sea quark for dynamical configurations)}
\label{tab:params}
\vspace*{10pt}
\begin{tabular}{lll|l}
\hline
$\kappa_{\rm sea}$ & 0.1375      &  0.1410 &$\infty$\\
\hline
$m_{PS}/m_V$       & 0.8048(9)   &  0.586(3) &--\\
$a^{-1}_\sigma$[GeV] & 0.937(6)    &  1.127(10) &0.919(7)\\
\#conf.    & 648         &  490    & 195\\
\hline
\end{tabular}
\vspace*{-6mm}
\end{table}

We take 5 $\kappa$ values for the light valence quark corresponding to 
$m_{\rm PS}/m_{\rm V}\approx 0.8-0.5$. 
The strange quark mass $m_s$ is fixed using the $K$ and the $\phi$ meson.  
Our results for the $B_s$ meson
are obtained with $m_s$ from the $K$, and the $\phi$ is used to
estimate the systematic error.

For the heavy quarks, we use NRQCD at $O(1/M)$ with a symmetric evolution 
equation as defined in~\cite{junko-colin}.  We employ 5 bare heavy quark 
masses, covering a range of roughly $2.5-4.5$ GeV. 

The heavy-light meson mass $M$ is determined from the difference of the 
meson energy at finite momentum and at rest, assuming the
dispersion relation, 
$E(\vec{p}) - E(0) = \sqrt{\vec{p}^2 + M^2} - M$.
As a consistency check, we use both the $B_d$ and the $B_s$ meson to
determine the b quark mass.  

In our calculation of decay constants, the heavy-light current is 
corrected through $O(\alpha/M)$. The mixing coefficients between the 
lattice operators~\cite{junko-colin} 
contributing at this order to the time component of the axial vector 
current $J_4$,  and the matching factor
to the continuum current has been calculated~\cite{kenichi} in one-loop
perturbation theory, 
\begin{eqnarray}
J_4 &=& (1 + \alpha \rho_0) J^{(0)}_{4,lat} 
 + (1 + \alpha \rho_1) J^{(1)}_{4,lat} \nonumber \\
 &+& \alpha \rho_2 J^{(2)}_{4,lat}.
\end{eqnarray}
 For the RG-improved gluon action, $\alpha_V$ has not
been calculated, and we use a tadpole-improved one-loop expression for 
the $\overline{MS}$ coupling, $\alpha^{TI}_{\overline{MS}}(1/a)$.
\section{Decay Constants}
\setlength\tabcolsep{0.3pc}
\begin{table}[t]
\caption{Results for decay constants. Errors given in this table are
statistical (including the statistical uncertainty in $M_b$), and, where
applicable, the uncertainty in fixing the strange quark mass. Other
systematic errors are discussed in the text.}
\label{tab:fB}
\vspace*{3mm}
\begin{tabular}{llll}
\hline
$\kappa_{\rm sea}$  & $f_B$[MeV]  & $f_{B_s}$[MeV] &  $ f_{B_s}/f_B$ \\
\hline
$\infty$            & 193(4)      & 221(4)(+7)      &  1.147(10)(35) \\
0.1375              & 216(4)      & 250(4)(+8)      & 1.157(9)(+35) \\
0.1410              & 215(6)      & 251(6)(+6)      &  1.166(14)(+31)  \\
\hline
\end{tabular}
\vspace*{-6mm}
\end{table}
\begin{figure}[t]
\vspace{-1cm}
\centerline{\hspace{1cm}
\epsfxsize=9.2cm
\epsfbox{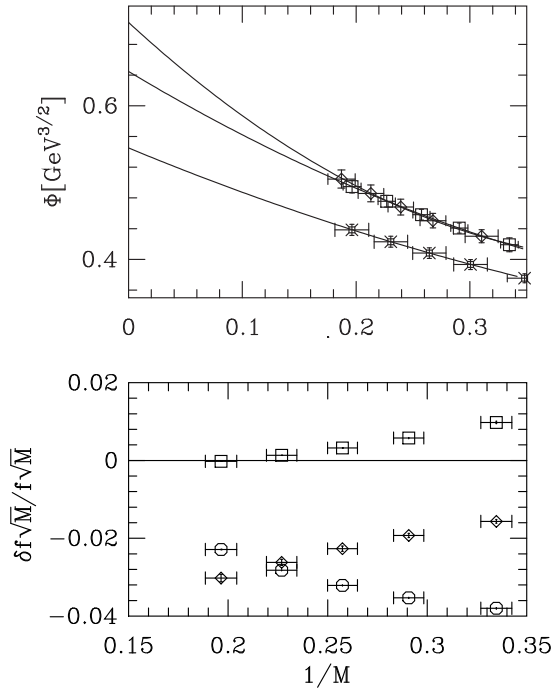}
}
\vspace{-2.4cm}
\caption{$\Phi \equiv (\alpha_s(M)/\alpha_s(M_B))^{(2/\beta_0)}
f\protect\sqrt{M}$ (top), and one-loop corrections to 
$f\protect\sqrt{M}$ (bottom) as a function of the
inverse pseudoscalar meson mass. In the upper plot,  squares stand
for $\kappa_{\rm sea} = 0.1375$, diamonds for $\kappa_{\rm sea} = 0.1410$
and fancy squares for quenched. In the lower plot, circles denote $\alpha
\rho_0 J^{(0)}_{4,lat}/J_4$,
squares, $\alpha \rho_1 J^{(1)}_{4,lat}/J_4$, and diamonds, $\alpha \rho_1 
J^{(2)}_{4,lat}/J_4$.
}
\vspace{-0.5cm}
\label{fig:fB}
\end{figure}
Our preliminary results for $f_B$, $f_{B_s}$ and $f_{B_s}/f_B$ are given  
in Table~\ref{tab:fB}, along with the statistical error  and, where
applicable, the uncertainty in the determination of $m_s$. 
Additional systematic errors are estimated as follows: $O(\alpha^2)$ 
corrections, taken to be $\alpha^2 \times O(1)$, are 5\%.  A previous
NRQCD calculation using the plaquette gluon action at $a^{-1} \sim 1$ GeV 
finds the tree level $O(1/M^2)$ corrections to be $\sim 2\%$~\cite{hein}; 
we estimate our error from the truncation of the $1/M$ expansion to be 
$\sim 4\%$.  The leading discretization effects from the light quarks 
and gluons of $O(\alpha a \Lambda_{QCD})$ and $O(a^2\Lambda_{QCD}^2)$ 
are 5\%. Added in quadrature, these estimates give $7\%$.

Our two-flavor results for $f_B$ and $f_{B_s}$ given in Table~\ref{tab:fB}
show a $10\%$ increase compared to the quenched values (see also 
Fig.~\ref{fig:fB}). We do not resolve any sea quark mass dependence. 
The dependence on the value of $\alpha_s$ is weaker than
for the plaquette gauge action, and the difference between renormalized and
bare decay constants  is only about $ 5\%$. 

In Fig.~\ref{fig:fB} we show
the one-loop corrections to the current $J_4$ as a function of the
heavy-light meson mass. In the $B$ region, $1/M \sim 0.2$, we find the
correction to $J^{(1)}_{4,lat}$ to be very small and the two other terms
to contribute about the same amount. The $J^{(2)}_{4,lat}$ contribution
also contains a discretization correction to the current first pointed out
in~\cite{junko-colin}. We note that this discretization correction is
considerably smaller for the RG gauge action than for the plaquette 
gauge action~\cite{kenichi}.

For $f_{B_s}/f_B$, we cannot resolve a difference between the three 
lattices. 

In a parallel study of B mesons using clover heavy quarks~\cite{hugh}, 
we have obtained $f_B$ and $f_{B_s}$ taking the chiral limit for 
sea quark at $\beta=1.8, 1.95$ and 2.1.  
The results from that study at $\beta=1.95$ agree within the 
estimated errors with the present results from NRQCD. 
\vspace{-10pt}
\section{Spectrum}
\begin{figure}[t]
\centerline{\hspace{1cm}
\epsfxsize=9.5cm
\epsfbox{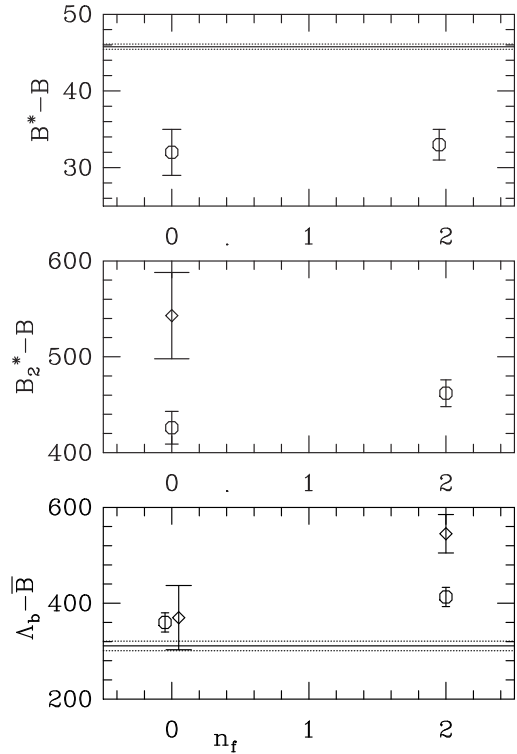}
}
\vspace{-2.4cm}
\caption{Meson and baryon splittings in MeV. 
Circles denote results from CP-PACS at
$a^{-1}(\protect\sqrt{\sigma}) \simeq 0.9$ GeV.  Diamonds stand for
results from~\protect\cite{lattice98} (quenched) and~\protect\cite{SGO}
($n_f$ = 2). Only statistical errors are shown. The solid line denotes
the experimental value, the dashed lines, its error.
}
\vspace{-0.4cm}
\label{fig:spectrum}
\end{figure}

%
%
In Fig.~\ref{fig:spectrum}, we  give our results for several $B$
splittings from the lattices with $n_f = 0$ and $n_f = 2, \kappa_{sea} = 
0.1375$. 
The top part of the figure shows the $B^\ast-B$ splitting. 
At present, we cannot resolve any unquenching effects. 
For quarkonia, on the 
same lattices, the hyperfine splitting is found to increase from the 
quenched value only by a few MeV~\cite{thomas}. 
We find the $B^\ast-B$ splitting to be $\sim 30 \%$ smaller than
the experimental value. Possible sources of systematic error are the
finiteness of the sea quark mass, the $O(\alpha)$ correction to the
coefficient of the $\sigma\cdot B$ operator, and higher order relativistic
corrections. 

%
In the middle part of Figure~\ref{fig:spectrum}, we show results for the
$B_2^\ast-B$ splitting, and in the lower part, the
spin-averaged $\Lambda_b-\overline{B}$ splitting. We do not find 
significant unquenching effects. However, for definite conclusions, we
need to study several sea quark masses and lattice spacings, which is in
progress.

%
%
%

\vspace{10pt}

This work is supported in part by the Grants-in-Aid
of Ministry of Education,
Science and Culture (Nos.~09304029, 10640246, 10640248, 10740107, 
11640250, 11640294, 11740162). 
SE and KN  are JSPS Research Fellows.
AAK and TM are supported by the Research for the Future 
Program of JSPS.

\end{document}